\documentclass[twocolumn,pre,final,showpacs,floatfix]{revtex4}
\usepackage{amsmath}
\usepackage{graphicx}
\usepackage{bm}
\usepackage{amssymb}
\usepackage{epstopdf}

\begin{document}

\title[Dimensional effects in ultrathin magnetic films]
{Dimensional effects in ultrathin magnetic films}

\author{Pavel V. Prudnikov, Vladimir V. Prudnikov, Maria A. Medvedeva}

\affiliation{Omsk State University,  pr. Mira 55, Omsk 644077 Russia}
\email{prudnikp@univer.omsk.su}
\begin{abstract}
Dimensional effects in the critical properties of multilayer Heisenberg films have been numerically studied by
Monte Carlo methods. The effect of anisotropy created by the crystal field of a substrate has been taken into
account for films with various thicknesses. The calculated critical exponents demonstrate a dimensional transition from two-dimensional to three-dimensional properties of the films with an increase in the number of
layers. A spin-orientation transition to a planar phase has been revealed in films with thicknesses corresponding to the crossover region.
\end{abstract}


\pacs{68.35.Rh, 68.55.jd, 75.40.Cx, 75.40.Mg}

\maketitle

\section{Introduction}
\label{sec:intro}

Great attention has been recently focused on
studying the properties of thin magnetic films \cite{Vaz_expFilms} primarily because the investigation of the physical properties of ferromagnetic films promotes the solution to
fundamental problems in the physics of magnetic phenomena and the development of the theory of ferromagnetism~\cite{bib:Bland_book}. Study of thin films significantly
expands concepts of the physical nature of the anisotropy of ferromagnets and makes it possible to reveal
diverse remagnetization processes and to observe new
physical phenomena \cite{2014_Ustinov_SPIN}. Structural states that can
hardly be obtained in bulk samples can be implemented in films. This significantly expands the possibilities of the study of the relation between the structural characteristics and physical properties of magnetic materials.

Study of the physical properties of thin ferromagnetic films is also topical for their technological applications. The most important application of films is
their use as magnetic media for writing and storage of
information in memory devices. Features of magnetic
films can promote an increase in the information writing density to 1 $\mathop{\mathrm{TBit/in^2}}$~\cite{HAMR_IEEE_2008}. In this respect, it is very
important to understand processes of variation of
magnetization in thin magnetic films with change in
the temperature, particularly at temperatures close to
the Curie temperature.

The critical properties of ultrathin films are sensitive to effects of anisotropy created by the critical field
of a substrate. These effects can be responsible for
change in the critical behavior of multilayer systems.

In this work, the physical properties of thin ferromagnetic films are numerically studied within the
anisotropic Heisenberg model with the Hamiltonian~\cite{BinderLandau}
\begin{equation}
H=-J\sum_{i,j}\left[(1-\Delta(N))\left(S_i^{\rm x} S_j^{\rm
x}+S_i^{\rm y} S_j^{\rm y}\right)+S_i^{\rm z} S_j^{\rm z}\right],
\end{equation}
where $\mathbf{S}_i=(S_i^{\rm x}, S_i^{\rm y},
S_i^{\rm z})$  is the three-dimensional unit
vector at the $i$th site, $J > 0$ characterizes the ferromagnetic exchange interaction between nearest spins, and $\Delta$
is the anisotropy parameter. The values $\Delta = 0$ and 1 correspond to the isotropic Heisenberg model and
Ising model, respectively.

\begin{figure}[b!]
\centering
 \includegraphics[width=0.4\textwidth]{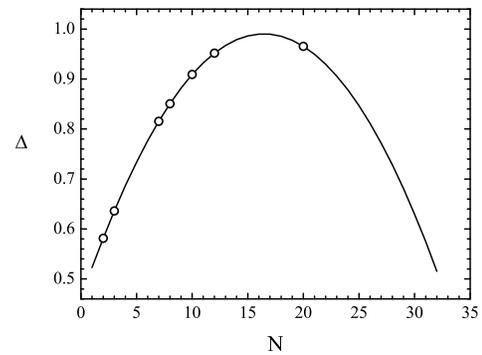}\\
\caption{\label{fig:DeltaN}Anisotropy parameter $\Delta(N)$ versus the thickness of
the film $N$. The circles are experimental data for $\mathrm{Ni(111)/W(110)}$
\cite{DimensionalCrossover_1992_Ni_W}. }
\end{figure}

The microscopic nature of anisotropy in Fe, Co,
and Ni films and its dependence on the thickness of a
film measured in the number of monoatomic layers $N$
are determined by the crystal field of the substrate, single-ion anisotropy, and dipole–dipole interaction
between the magnetic moments of atoms in the film
and their mutual competition. For this reason, the calculation of the anisotropy effects in magnetic films is
very difficult. The effective dependence of the anisotropy parameter $\Delta(N)$ on the thickness of the film $N$ is
chosen proportional to the $N$ dependence of the critical temperature for $\mathrm{
Ni(111)/W(110)}$ films \cite{DimensionalCrossover_1992_Ni_W} with different numbers of layers. In the approximation procedure for the dependence $\Delta(N)$, we used the fact that $\mathrm{Ni}$
films with a large number of layers exhibit bulk critical
properties corresponding to three-dimensional isotropic Heisenberg magnets \cite{OrientalOrderParam_1995_PRL,OrientalOrderParam_2013}.
The resulting dependence $\Delta(N)$ is shown in Fig.~\ref{fig:DeltaN}.

The Monte Carlo simulation was performed for
$L \times L \times N$ films with periodic boundary conditions in
the plane of the film. The number of spins in each layer
is $L\times L$ and the number of layers in the thin film is $N$.
In this work, we considered systems with linear dimensions $L = 32$, $48$, and $64$ and the number of layers ranging from $N = 1$ to $32$. The characteristics of the film
were calculated for the temperature interval $T = 0.01 - 5.01$ with a step $\Delta T_{\mathrm{step}} = 0.02$.

\begin{figure*}[t!]
\begin{center}
\includegraphics[width=0.4\textwidth]{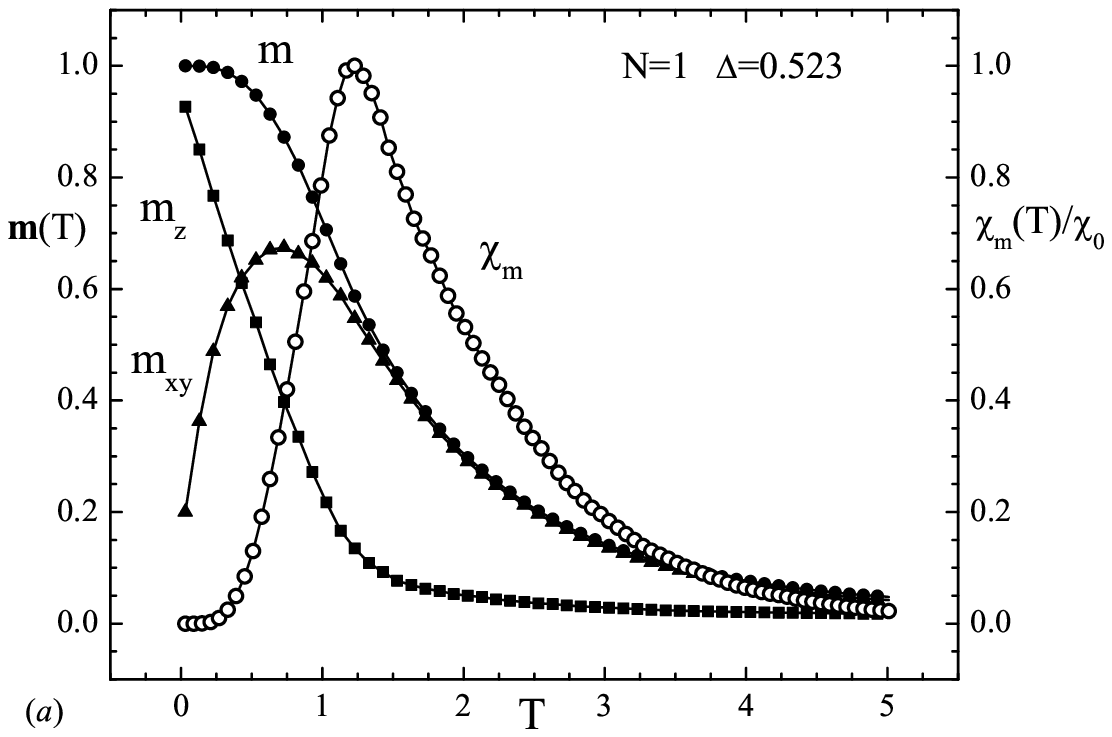}
\includegraphics[width=0.4\textwidth]{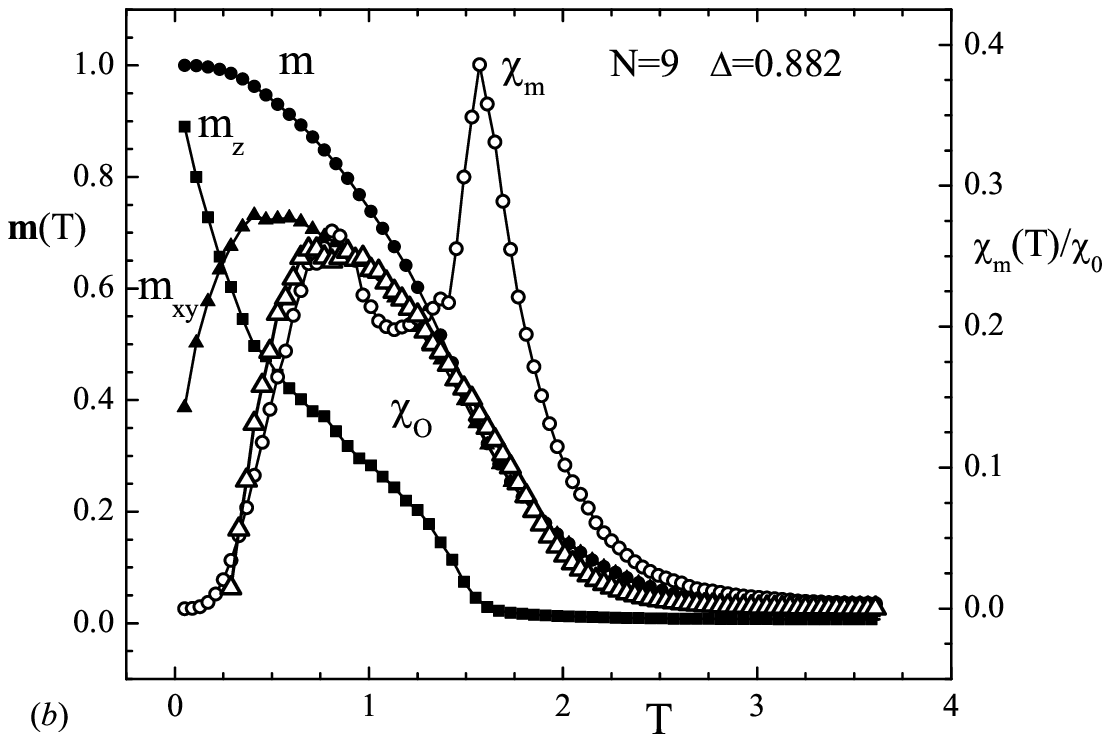}\\
\caption{\label{fig:N} Temperature dependencies of the magnetizations
$m(T)$, $m_z(T)$, and $m_{xy}(T)$ and susceptibilities $\chi_m(T)$, $\chi_O(T)$
for the number of layers $N = 1$ (a) and 9 (b).}
\end{center}
\end{figure*}

Within the three-dimensional anisotropic Heisenberg model with the use of the Swendsen–Wang algorithm, we studied the temperature dependencies of the
magnetization
\begin{equation}\label{eq:m}
m= \Biggl\langle \frac{1}{N_s} \biggl[ \sum\limits_{\alpha \in
\{x,y,z\}}
  \left(\textstyle\sum\nolimits_{i}^{N_s} S_i^{\alpha}\right)^2
  \biggr]^{1/2} \Biggr\rangle,
\end{equation}
its components, i.e., the magnetization normal to the
plane of the film
\begin{equation}\label{eq:mz}
m_z= \biggl\langle \frac{1}{N_s}  \textstyle\sum\nolimits_{i}^{N_s}
S_i^{\rm z} \biggr\rangle,
\end{equation}
and the magnetization in the plane of the film
\begin{equation}\label{eq:mxy}
m_{\rm xy}= \biggl\langle \frac{1}{N_s}  \biggl[
\left(\textstyle\sum\nolimits_{i}^{N_s} S_i^{\rm x}\right)^2 +
\left(\textstyle\sum\nolimits_{i}^{N_s} S_i^{\rm y}\right)^2
\biggr]^{1/2} \biggr\rangle,
\end{equation}
as well as the orientational order parameter~\cite{{OrientalOrderParam_1995_PRL},{OrientalOrderParam_2013}}
\begin{equation}\label{eq:OP_orient}
O_{\alpha}=\left\langle\
\displaystyle\left|\frac{n^\alpha_h-n^\alpha_v}{n^\alpha_h+n^\alpha_v}\right|
\ \right\rangle,
\end{equation}
Here, $N_s = NL^2$ is the total number of spins in the film,
angular brackets mean statistical averaging, $\alpha \in \{x,y,z\}$, and $n_h$ and $n_v$ are the numbers of the horizontal and
vertical pairs of nearest spins with oppositely directed
$S_z$, respectively:
\begin{eqnarray}
n^\alpha_h&=&\sum_{\mathbf{r}}\biggl\{ 1- \mathop{\mathrm{sgn}}\left[ S^{\alpha}(r_x,r_y),S^{\alpha}(r_x+1,r_y) \right]  \biggr\}, \nonumber \\
n^\alpha_v&=&\sum_{\mathbf{r}}\biggl\{ 1-
\mathop{\mathrm{sgn}}\left[
S^{\alpha}(r_x,r_y),S^{\alpha}(r_x,r_y+1) \right]  \biggr\}.
\end{eqnarray}

The critical behavior of the system near the phase-transition temperature is clearly characterized by the
magnetic susceptibility
\begin{equation}
\chi_m \sim [\langle m^2 \rangle]-[\langle m \rangle]^2.
\end{equation}

The dependence $\chi_m(T)$ characterizes critical fluctuations of the magnetization and their interaction. The
temperature at the maximum of the temperature
dependence $\chi_m(T)$ can be used to estimate the temperature of a ferromagnetic phase transition in the film
and its dimensional changes for various thicknesses of
the film.

The temperature behavior of the orientational susceptibility
\begin{equation}
\chi_o \sim [\langle O_{\alpha}^2 \rangle]-[\langle O_{\alpha}
\rangle]^2.
\end{equation}

makes it possible to reveal the region of the spin-orientation transition from a phase in which the magnetization is normal to the plane of the film to a phase where
the magnetization preferentially orients in the plane of
the film.

To more accurately determine the critical temperature of the transition from the paramagnetic phase to
the ferromagnetic one, we found the temperature
dependence of the Binder cumulant:
\begin{equation}\label{eq:U4}
U_4(T,L) = \frac{1}{2}\left( 3 -
\frac{[\left<m^4(T,L)\right>]\phantom{^2}}{[\left<m^2(T,L)\right>]^2}\right).
\end{equation}

The scaling dependence of the cumulant
\begin{equation}\label{eq:U4scail}
U_4(T,L) = u\left(L^{1/\nu}(T-T_c)\right).
\end{equation}
allows determining the temperature of a second-order
phase transition from the coordinate of the intersection of the temperature dependencies $U_4(T, L)$ for different $L$ values.

In this work, we considered the finite-dimensional
scaling form \cite{bib:DillmannJanke} in films for the quantities
\begin{eqnarray}
\label{eq:scaling}
\langle m(T,N)\rangle &=& L^{-\beta/\nu}\tilde{m}(L^{1/\nu}\tau, N)\label{eq:chi_N}\\
\chi_m(T,N) &=& L^{\gamma/\nu}\tilde{\chi}(L^{1/\nu}\tau,
N)\nonumber
\end{eqnarray}
where $\gamma$, $\beta$, and $\nu$ are the effective critical exponents of
the susceptibility $\chi_m$, magnetization $m$, and correlation length $\xi$, respectively. Scaling form (\ref{eq:chi_N}) determines the dependence of $m$ and $\chi_m$ on the linear
dimension $L$ and the number of the layers $N$ in the film
and makes it possible to determine the effective critical
exponents from the resulting temperature dependencies of these quantities.
The detailed analysis of the behavior of the magnetizations $m$, $m_{xy}$, and $m_z$ and the susceptibilities $\chi_m$ and
$\chi_O$ for films with various thicknesses revealed two types
of phase transitions. A transition from the ferromagnetic phase to the paramagnetic one is observed for all
$N$ values under consideration. The indicated transition
is accompanied by a peak of the magnetic susceptibility $\chi_m$. These magnetizations and susceptibilities for
films with $N = 1$ and $9$ are shown in Fig.~\ref{fig:N}. According
to the data shown in Fig.~\ref{fig:N}a, although $m_z$ vanishes in
the region of the phase transition temperature, the
critical nature of the transition is determined by fluctuations of $m_z$. This statement for a monolayer film is
confirmed by the results reported in \cite{BinderLandau}. The critical
temperatures of the ferromagnetic phase transition
were determined more accurately by the method of
intersection of Binder cumulants (\ref{eq:U4}).

For the interval of the thicknesses of films $N = 9 - 22$, an additional peak in the high-temperature region
appears in the dependence $\chi_m(T)$ owing to the spin-orientation transition whose nature is confirmed by
the behavior of the orientational susceptibility $\chi_O$. The
behavior of the magnetizations $m$, $m_{xy}$, and $m_z$ and the
susceptibilities $\chi_m$ and $\chi_O$ for $N = 9$ is shown in Fig.~\ref{fig:N}b.

The specification of the critical behavior of the film
to a certain universality class, as well as the effective
dimension of the system, can be characterized by a set
of critical exponents. The critical exponent $\nu$ can be
determined from the scaling behavior of Binder cumulants (\ref{eq:U4}), and the critical exponents $\beta/\nu$ and $\gamma/\nu$ can
be found from scaling dependencies (\ref{eq:scaling}) of the magnetization $m$ and susceptibility $\chi$ on $L$ at the corresponding critical temperature $T_c$.

\begin{figure}[t!]
\begin{center}
\includegraphics[width=0.4\textwidth]{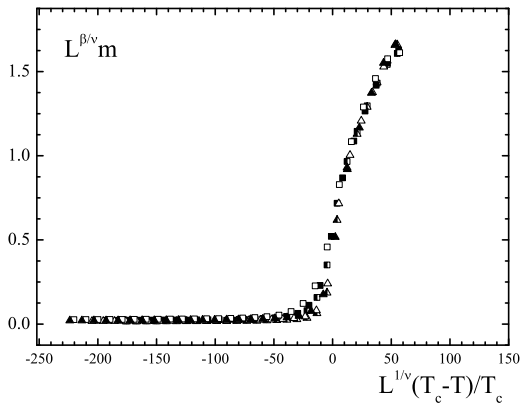}
\caption{\label{fig:skeil}Scaling dependence $f(m) =
L^{\beta/\nu}m$ versus $L^{1/\nu}(T_c-T)/T_c$.}
\end{center}
\end{figure}

The scaling dependencies $f(m) = L^{\beta/\nu}m$ on $L^{1/\nu}(T_c - T)/T_c$ calculated for films with the linear dimensions
$L = 32$, $48$, and $64$ and thicknesses from $2$ to $5$ ML
exhibit "collapse" of the data (Fig.~\ref{fig:skeil}) on a single universal curve. This confirms that the effective critical
exponents are calculated correctly.

\begin{figure}[t!]
\centering
 \includegraphics[width=0.4\textwidth]{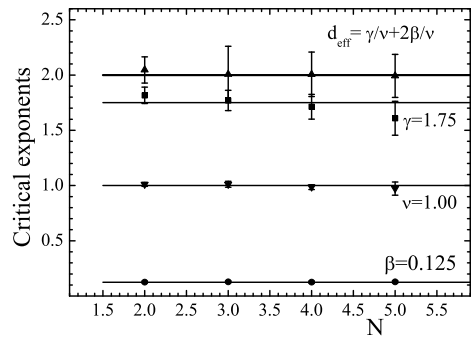}\\
\caption{\label{fig:CrExponents}Critical exponents $\nu$, $\beta$, $\gamma$ and effective dimension $d_{\mathrm{eff}}$.}
\end{figure}

The analysis of the temperature dependencies of the
model quantities for various numbers of layers in the
film $N$ shows that they are separated into several
groups with different asymptotic critical behaviors. In
particular, films with the number of layers $N < 5$ exhibit
a behavior with critical exponents close to the exact
values for the two-dimensional Ising model ($\beta_{\mathrm{2D Ising}} =
1/8$, $\nu_{\mathrm{2D Ising}} = 1$, $\gamma_{\mathrm{2D Ising}} = 7/4$; Fig.~4): $\beta = 0.126(8)$,
$\nu = 1.010(17)$, and $\gamma = 1.816(69)$ at $N = 2$; $\beta =
0.128(8)$, $\nu = 1.011(27)$, and $\gamma = 1.770(94)$ at $N = 3$;
$\beta = 0.126(9)$, $\nu = 0.986(21)$, and $\gamma = 1.713(112)$ at
$N = 4$; and $\beta = 0.129(9)$, $\nu = 0.972(59)$, and $\gamma =
1.609(150)$ at $N = 5$.
Using the hyperscaling relation $\gamma/\nu + 2\beta/\nu = d$, the
effective dimension of the system $d_{\mathrm{eff}}$ can be obtained.
For films with the thickness $N = 2 - 5$, $d_{\mathrm{eff}}$ is close to 2
(see Fig.~4). Thus, the films with $N/L \ll 1$ exhibit the
critical behavior characteristic of quasi-two-dimensional systems.
The resulting critical exponent $\beta(N)$ presented in
Fig.~5 clearly exhibits the transition from the behavior
of the two-dimensional Ising model to the three-dimensional Heisenberg model with an increase in the
thickness of the film. The dimensional transition demonstrated in this work from two-dimensional to three-dimensional critical properties of multilayer magnetics with an increase in the thickness of the film is in good agreement with the experimental data (see Fig.~2
in \cite{DimensionalCrossover_1992_Ni_W} and Fig.~7 in \cite{Vaz_expFilms}).

\begin{figure}[t!]
\centering
\includegraphics[width=0.4\textwidth]{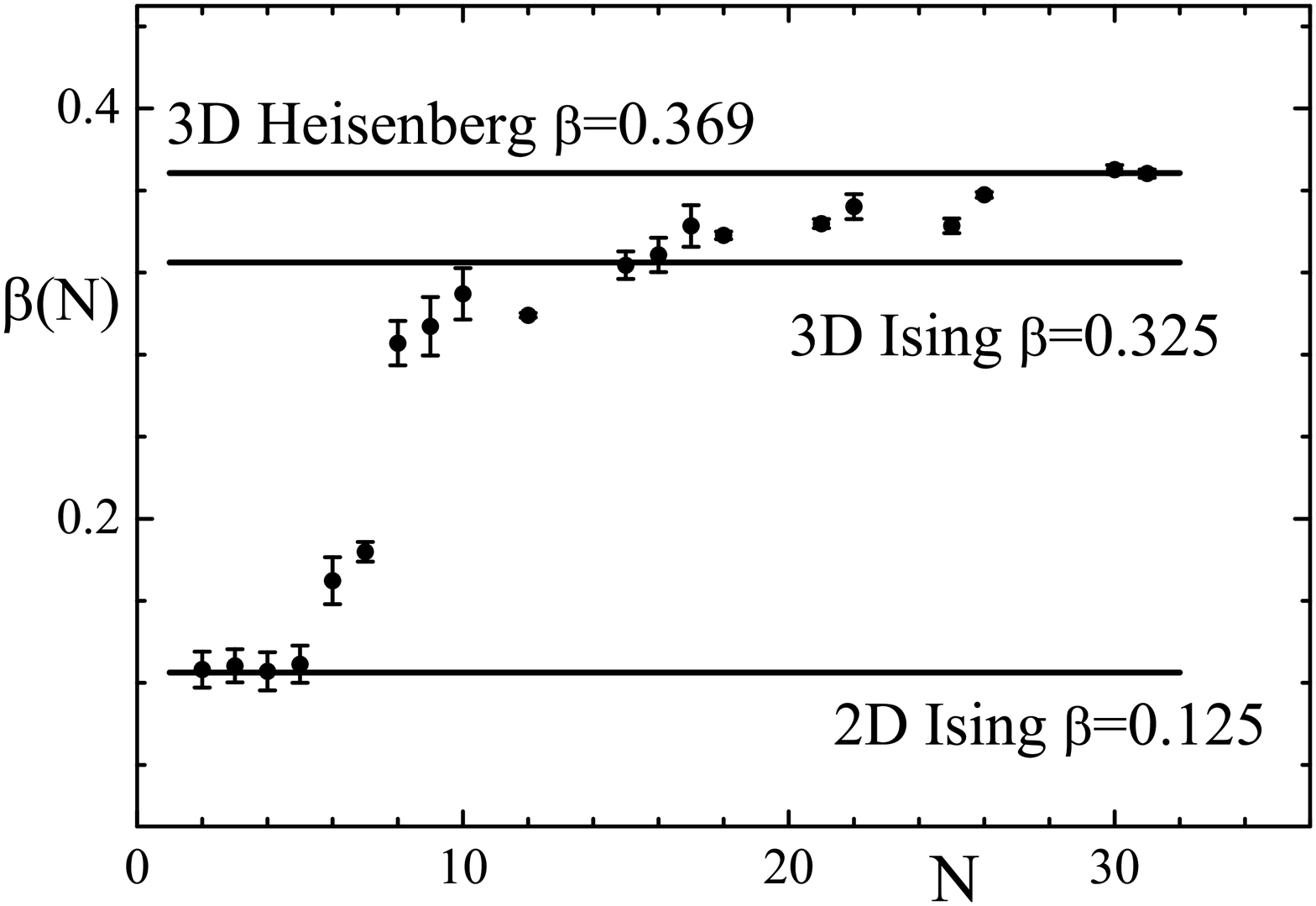}\\
\caption{\label{fig:beta_N}Critical exponent $\beta$ versus the thickness of the film $N$.}
\end{figure}

Figure 6 shows the resulting phase diagram of
change in the state of thin-film systems with an
increase in their thickness. The systems with $N \leq 5$
exhibit the critical behavior corresponding to the two-dimensional Ising model. The critical behavior of
films with $6 \leq N \leq 12$ corresponds to the crossover
region of the transition from two-dimensional properties to three-dimensional ones. The critical characteristics for films with the thicknesses $N = 13 - 21$ correspond to the three-dimensional Ising model. In films
with $9 \leq N \leq 22$, the spin-reorientation transition occurs
to the planar phase with the critical behavior close to
the behavior of the XY model. Since the planar phase
is revealed, it is necessary to significantly expand the
necessary set of data at the statistical averaging of the
calculated quantities. Films with $N \geq 22$ exhibit the
bulk critical behavior corresponding to an isotropic
Heisenberg magnet. The results are confirmed by the
critical exponents presented in the table.

\begin{figure}[t!]
\begin{center}
\includegraphics[width=0.4\textwidth]{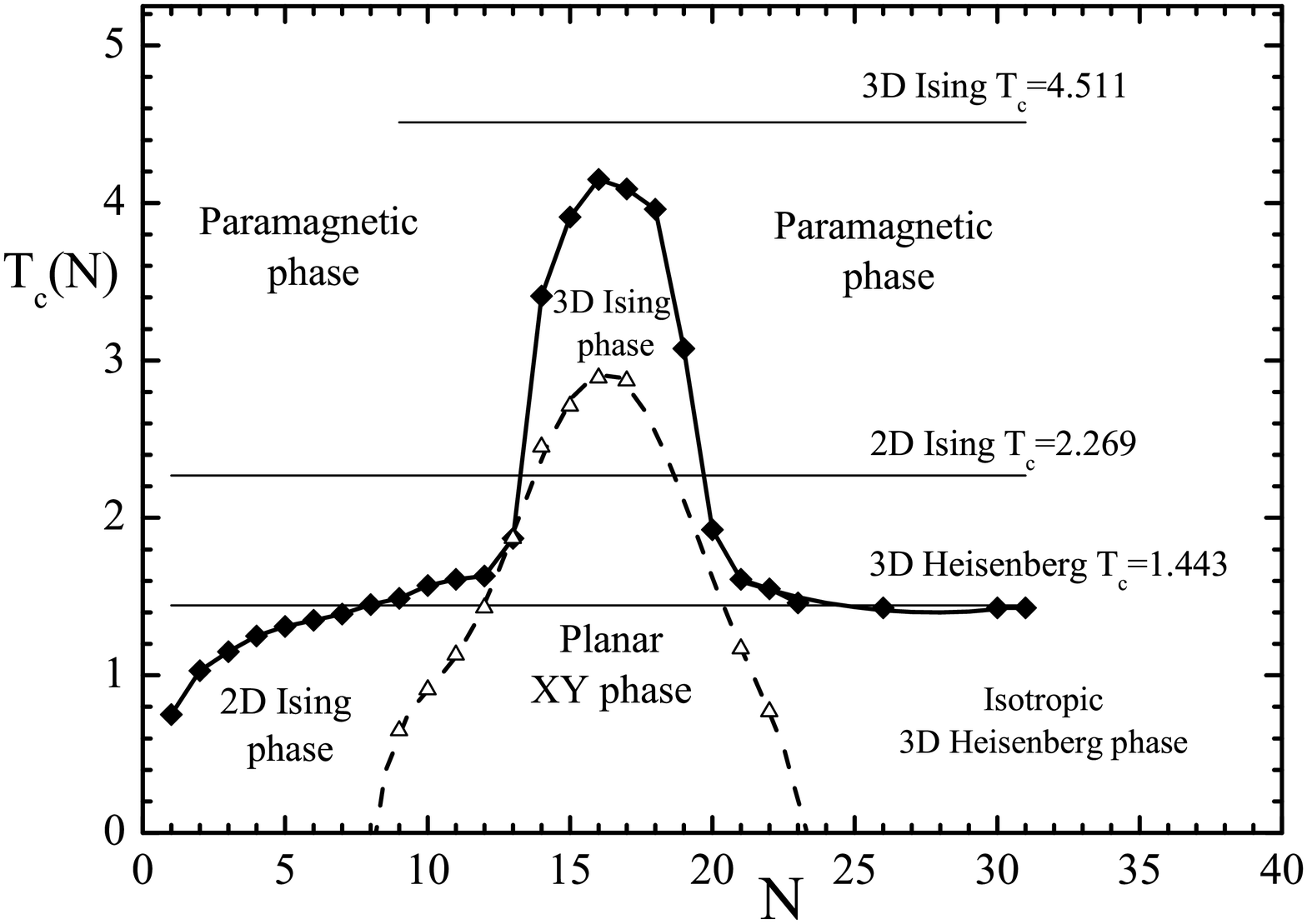}\\
\end{center}
\caption{\label{fig:phase_diarg}Phase diagram for thin films. The solid line corresponds to the transition from the ferromagnetic phase to
the paramagnetic phase and the dashed line corresponds to
the spin-reorientation transition.}
\end{figure}

\begin{table}[!t]
\caption{Critical exponents $\beta$, $\nu$ and
$\gamma$ }\label{tab:CrExponents}
\begin{center} \small \tabcolsep 2pt
\begin{tabular}{llll}
\hline
 & \multicolumn{1}{c}{$\beta$} & \multicolumn{1}{c}{$\nu$} & \multicolumn{1}{c}{$\gamma$} \\
\hline\hline
$N=2 \div 5$            & 0.127(2)  & 0.995(10) & 1.727(45)  \\
\hline
2D Ising model   & 0.125     & 1.0       & 1.750      \\
\hline
$N=6$              & 0.170(11) & 0.974(62) & 1.803(76)  \\
$\phantom{N=}\ 8$  & 0.286(11) & 0.981(28) & 1.586(52)  \\
$\phantom{N=}\ 10$ & 0.310(13) & 0.839(32) & 1.225(61)  \\
\hline
$N=15$             & 0.324(7)  & 0.632(21) & 1.141(33)  \\
$\phantom{N=}\ 16$ & 0.329(8)  & 0.634(28) & 1.182(53)  \\
$\phantom{N=}\ 17$ & 0.343(10) & 0.658(27) & 1.132(42)  \\
\hline 3D Ising model \cite{2000_3DIsing_JankeVillanova}
                   & 0.3249(6) & 0.6299(5) & 1.2331(13) \\
\hline
$N=26$             & 0.358(1)  & 0.723(18) & 1.396(121) \\
$\phantom{N=}\ 31$   & 0.368(2)  & 0.759(40) & 1.414(77)  \\
\hline 3D Heisenberg model \cite{2000_3DHeis_Hasen_JPA}
                   & 0.3685(11)& 0.710(2)  & 1.393(4)  \\
\hline
\end{tabular}
\end{center}
\end{table}

To summarize, the numerical investigation of the
magnetic properties and critical behavior of thin films
within the anisotropic Heisenberg model has revealed
dimensional effects in the behavior of the magnetization and magnetic susceptibility. A spin-reorientation
transition has been identified in films with the thicknesses $N = 9 - 22$ ML. The critical exponents $\beta$, $\nu$, and
$\gamma$ have been calculated for films with various thicknesses. The resulting averaged critical exponent $\gamma =
1.73(5)$ for $N \leq 5$ is in good agreement with the experimental value $\gamma = 1.75(2)$ measured for a $\mathrm{\mathop{Fe/W(110)}}$
bilayer film in \cite{bib:2004_FeW_gamma}. A transition from the two-dimensional to three-dimensional properties in the behavior
of multilayer magnets with an increase in the thickness
of the film has been revealed for the first time.

We are grateful to L.N. Shchur, A.K. Murtazaev,
and M.V. Mamonova for discussion of the results. This
work was supported by the Russian Science Foundation (project no. 14-12-00562). For our calculations,
we used the resources provided by the supercomputer
center of the Moscow State University and Joint Supercomputer Center of the Russian Academy of Sciences.


\begin{thebibliography}{10}

\bibitem{Vaz_expFilms}
C.~A.~F.~Vaz, J.~A.~C.~Bland, and G.~Lauhoff, Rep. Prog. Phys. {\bf
71}, 056501 (2008).

\bibitem{bib:Bland_book}
J.\,A.\,C. Bland  and B.\,Heinrich, {\sl Ultrathin Magnetic
Structures} IV: Berlin: Springer, 2005 - 257p.

\bibitem{2014_Ustinov_SPIN}
V.\,V.~Ustinov, M.\,A. Milyaev, and L.\,I.
Naumova, SPIN {\bf 04}, 1440001 (2014).

\bibitem{HAMR_IEEE_2008}
M.\,A. Seigler, W.\,A. Challener, E. Gage, N. Gokemeijer, G. Ju, B.
Lu, K. Pelhos, C. Peng, R. E. Rottmayer, X. Yang, H. Zhou, and T.
Rausch IEEE Trans. Magn. {\bf 44}, 119 (2008).

\bibitem{BinderLandau}
K. Binder, D.\,P. Landau, Phys. Rev. B. {\bf 13}, 1140 (1976).

\bibitem{DimensionalCrossover_1992_Ni_W}
Y. Li and K. Baberschke, Phys. Rev. Lett. {\bf 68}, 1208 (1992).

\bibitem{OrientalOrderParam_1995_PRL}
I. Booth, A.\,B. MacIsaac, J.\,P. Whitehead, K.~De'Bell, Phys. Rev.
Lett. {\bf 75}, 950 (1995)

\bibitem{OrientalOrderParam_2013}
M.\,C. Ambrose and R.\,L. Stamps, Phys. Rev. B. {\bf 87}, 184417 (2013).

\bibitem{bib:DillmannJanke}
O. Dillmann, W. Janke, M. Muller, and K. Binder, J. Chem. Phys. {\bf
114}, 5853 (2001).

\bibitem{1994_Ni_films}
F.~Huang, M.~T.~Kief, G.~J.~Mankey, and R.~F.~Willis, Phys. Rev. B
{\bf 49}, 3962 (1994).

\bibitem{2001_Ni_films}
A.~Ney, A.~Scherz, P.~Poulopoulos, K.~Lenz, H.~Wende, K.~Baberschke,
F.~Wilhelm and N.\,B.~Brookes Phys. Rev. B {\bf 65}, 024411 (2001).

\bibitem{2000_3DIsing_JankeVillanova}
W.~Janke, D.A.~Johnston, R.~Villanova, Physica A {\bf 281}, 207
(2000).

\bibitem{2000_3DHeis_Hasen_JPA}
M.~Hasenbush, J. Phys. A {\bf 34}, 8221 (2001).

\bibitem{bib:2004_FeW_gamma}
M.\,J. Dunlavy and D. Venus, Phys. Rev. B. {\bf 69},
094411 (2004).

\end{thebibliography}
\end{document}